\documentclass{article}
\usepackage{ijcai22}
\usepackage{amsmath}
\usepackage{amsthm}
\usepackage{amssymb}
\usepackage[english]{babel}
\usepackage{proof}
\usepackage{xspace}
\usepackage{esvect} 

\usepackage{ebproof}	
\usepackage{graphicx}   

\usepackage{cancel}
\usepackage{xcolor}

\usepackage{wasysym} 
 
\usepackage{tikz}
\usetikzlibrary{arrows, positioning, calc}
\usetikzlibrary{automata,positioning}

\usepackage{multicol}
\usepackage{longtable}

\usepackage{cancel} 

\usepackage{comment}

\usepackage{xcolor}

\usepackage[hidelinks]{hyperref} 

\usepackage{centernot} 

\usepackage{grid-system}

\usepackage{tabularx}

\usepackage{listings} 
\newtheorem{theorem}{Theorem}[section]

\newtheorem{lemma}[theorem]{Lemma}

\theoremstyle{definition}
\newtheorem{definition}[theorem]{Definition}

\newtheorem{example}[theorem]{Example}

\newcommand{\I}{\emph{I}\xspace}
\newcommand{\You}{\emph{You}\xspace}
\newcommand{\My}{\emph{My}\xspace}

\newcommand{\Me}{\emph{Me}\xspace}
\newcommand{\Your}{\emph{Your}\xspace}
\newcommand{\Yours}{\emph{Yours}\xspace}
\newcommand{\x}{\vv{\times}}

\newcommand{\dI}{\mathrm{deg}_\mathcal{I}}

\graphicspath{ {./images/} }

\usepackage[backend=biber,style=alphabetic,]{biblatex}
\usepackage{times}

\usepackage{soul}
\usepackage{url}
\usepackage[utf8]{inputenc}
\usepackage[small]{caption}
\usepackage{graphicx}
\usepackage{amsmath}
\usepackage{booktabs}
\usepackage{todonotes}
\title {Truth and Preferences - A Game Approach for Qualitative Choice Logic\footnote{Paper accepted to M-pref 2022}}

\author{
Robert Freiman$^1$\and
Michael Bernreiter$^1$\\
\affiliations
$^1$Institute of Logic and Computation, TU Wien\\
\emails
robert@logic.at,
michael.bernreiter@tuwien.ac.at
}

\begin{document}
	\maketitle
	
\begin{abstract}
    In this paper, we introduce game-theoretic semantics (GTS) for Qualitative Choice Logic (QCL), which, in order to express preferences, extends classical propositional logic with an additional connective called ordered disjunction. Firstly, we demonstrate that game semantics can capture existing degree-based semantics for QCL in a natural way. Secondly, we show that game semantics can be leveraged to derive new semantics for the language of QCL. In particular, we present a new semantics that makes use of GTS negation and, by doing so, avoids problems with negation in existing QCL-semantics.
\end{abstract}

\section{Introduction}
Preferences are a key research area in artificial intelligence, and thus a multitude of preference formalisms have been described in the literature \cite{PigozziEtAl2016preferences}. An interesting example is Qualitative Choice Logic (QCL) \cite{BrewkaEtAl2004qcl}, which extends classical propositional logic by the connective $\x$ called ordered disjunction. Intuitively, $F \x G$ states that $F$ or $G$ should be satisfied, but satisfying $F$ is more preferable than satisfying only $G$. 
This allows to express soft-constraints (preferences) and hard-constraints (truth) in a single language. 

For example, let us say we want to formalize our choice of pizza toppings, and that we definitely want tomato-sauce~($t$). Moreover, we want either mushrooms~($m$) or artichokes~($a$), but preferably mushrooms. This can easily be expressed in QCL via the formula $t \land (m \x a)$.
This formula has three models in QCL, namely  
$M_1 = \{t,\allowbreak m,\allowbreak a\}$,
$M_2 = \{t,\allowbreak m\}$, and 
$M_3 = \{t,\allowbreak a\}$. QCL-semantics then ranks these models via so-called satisfaction degrees. The lower this degree, the more preferable the model. In this case, $M_1$ and $M_2$ would be assigned a degree of $1$ and $M_3$ would be assigned a degree of $2$, i.e., $M_1$ and $M_2$ are the preferred models of this formula. 

In the literature, QCL has been studied with regards to possible applications \cite{BrewkaEtAl2004lpods, BenferhatSedki2008qclalert, LietardEtAl2014qcldatabase} and computational properties \cite{BernreiterEtAl2021}. 
However, not all aspects of QCL-semantics are uncontroversial. For example, viewing QCL as an extension of classical logic, it is natural to expect a formula $F$ to be logically equivalent to its double negation $\neg \neg F$. But this property does not hold in QCL, as all information about preferences in $F$ is is erased in $\neg F$. This issue has been addressed by Prioritized QCL (PQCL) \cite{BenferhatSedki2008pqcl}, which defines ordered disjunction in the same way as QCL but changes the meaning of the classical connectives, including negation. While PQCL solves QCL's problem with double negation, it in turn introduces other controversial behavior, e.g., that both a formula $F$ and its negation $\neg F$ can be satisfied by the same interpretation. No alternative semantics for QCL is known to us that addresses both of these issues at the same time. 

In order to tackle these issues, we develop game-theoretic semantics (GTS) for QCL, embedding choice logics in the rich intersection of the fields of game-theory and logics (\cite{sep-logic-games,vbenthbook,vbook}). Building on the concepts of rational behavior and strategic thinking, GTS offer a natural dynamic viewpoint of dealing with truth and preferences. Originally, GTS go back to Jaakko Hintikka \cite{Hintikka1973-HINLLA}, who designed a win/lose game for two players, called \Me (or \I) and \You\footnote{Hintikka and others call this player \emph{Nature}}, both of which can act in the role of Proponent or Opponent of a formula $F$ of over an interpretation $\mathcal{I}$. The game proceeds by rules for step-wise reducing $F$ to an atomic formula. Most importantly, negation is interpreted in game-theoretic terms as \emph{dual negation}, \cite{tulenheimo}: at formulas $\neg G$, the game continues with $G$ and a role switch. It turns out that \I have a winning strategy for this game if and only if $F$ is classically true over $\mathcal{I}$.

To capture not only truth but also preferences, we extend this two-valued game with more fine-grained outcomes. We show that our proposed game framework adequately models the degree-semantics of QCL. Game-theoretically speaking, the aforementioned issues with negation in QCL arise, because the players in the GTS are not fully symmetric. Eliminating this asymmetry leads to a new game and by extension yields a new logic we call Game-induced Choice Logics (GCL), where negation indeed behaves as in classical logic. In the last section, we outline how to lift the GTS to a sequent calculus for preferred model entailment in GCL.

\section{Preliminaries}

In this section, we formally introduce QCL and discuss fundamental notions in GTS.

\subsection{Qualitative Choice Logic (QCL)}

The most prominent choice logic in the literature is QCL \cite{BrewkaEtAl2004qcl}, which adds ordered disjunction ($\x$) to classical propositional logic. 

\begin{definition}
    Let $\mathcal{U}$ denote an infinite set of propositional variables. The set $\mathcal{F}_{\mathit{QCL}}$ of QCL-formulas are built inductively as follows:
	(i) $a \in \mathcal{F}_{\mathit{QCL}}$ for all $a \in \mathcal{U}$;
	(ii) if $F \in \mathcal{F}_{\mathit{QCL}}$, then $(\neg F) \in \mathcal{F}_{\mathit{QCL}}$;
	(iii) if $F, G \in \mathcal{F}_{\mathit{QCL}}$, then $(F \circ G) \in \mathcal{F}_{\mathit{QCL}}$ for $\circ \in \{\land, \lor, \x\}$. 
\end{definition}


The semantics of QCL is based on two functions, namely optionality and satisfaction degree. The satisfaction degree of a formula can be either a natural number or $\infty$ and is used to rank interpretations: the lower the degree, the better. The optionality of a formula represents the maximum finite satisfaction degree this formula can obtain (as we will see in Lemma~\ref{degopt}) and is used to penalize interpretations that do not satisfy the preferred option $F$ in an ordered disjunct $F \x G$.

\begin{definition}\label{def:satisfaction-degree}
    The optionality of QCL-formulas is defined inductively as follows: 
    (i)~$\mathrm{opt}(a) = 1$ for every propositional variable $a \in \mathcal{U}$,
    (ii)~$\mathrm{opt}(\neg F) = 1$,
    (iii)~$\mathrm{opt}({F \circ G}) = \max(\mathrm{opt}(F),\mathrm{opt}(G))$ for  $\circ \in \{\vee,\wedge\}$, and
    (iv)~$\mathrm{opt}({F\x G}) = \mathrm{opt}(F) + \mathrm{opt}(G)$.
\end{definition}

\begin{definition}
    An \emph{interpretation} $\mathcal{I} \subseteq \mathcal{U}$ is a set of propositional variables. The satisfaction degree of QCL-formulas is defined inductively as follows:
    \begin{align*}
        \mathrm{deg}_\mathcal{I}(a) &= 
            1 \text{ if } a \in \mathcal{I}, \infty \text{ otherwise } \\
        \mathrm{deg}_\mathcal{I}(\neg F) &= 
            1 \text{ if } \mathrm{deg}_\mathcal{I}(F) = \infty, \infty \text{ otherwise } \\
        \mathrm{deg}_\mathcal{I}(F\wedge G) &= \max(\mathrm{deg}_\mathcal{I}(F),\mathrm{deg}_\mathcal{I}(G))\\
        \mathrm{deg}_\mathcal{I}(F\vee G) &= \min(\mathrm{deg}_\mathcal{I}(F),\mathrm{deg}_\mathcal{I}(G)) \\
        \mathrm{deg}_\mathcal{I}(F\x G) &= 
            \begin{cases}
            	\mathrm{deg}_\mathcal{I}(F) & \text{if } \mathrm{deg}_\mathcal{I}(F) < \infty \\
                \mathrm{opt}(F) + \mathrm{deg}_\mathcal{I}(G) & \text{if } \mathrm{deg}_\mathcal{I}(F) = \infty, \\ & \dI(G) < \infty\\
                \infty &\text{otherwise}
            \end{cases}
    \end{align*}
\end{definition}

If $\mathrm{deg}_\mathcal{I}(F) = k$ we say that $\mathcal{I}$ satisfies $F$ to a degree of~$k$. If $\mathrm{deg}_\mathcal{I}(F) < \infty$ we say that $\mathcal{I}$ classically satisfies $F$, or that $\mathcal{I}$ is a model of $F$. Moreover, if $\mathrm{deg}_\mathcal{I}(F) < \mathrm{deg}_\mathcal{J}(F)$ for two interpretations $\mathcal{I}$ and $\mathcal{J}$ then $\mathcal{I}$ is more preferable than $\mathcal{J}$. 
To fully understand QCL-semantics, we must take note that satisfaction degrees are bounded by optionality, as intended:
\begin{lemma}[from \cite{BrewkaEtAl2004qcl}]\label{degopt}
For all QCL-formulas $F$ and all interpretations $\mathcal{I}$, $\mathrm{deg}_\mathcal{I}(F) \leq \mathrm{opt}(F)$ or $\mathrm{deg}_\mathcal{I}(F) = \infty$.
\end{lemma}

Indeed, inspecting Definition~\ref{def:satisfaction-degree} in view of Lemma~\ref{degopt} shows how optionality is used to penalize non-satisfaction: given $F \x G$, if some interpretation $\mathcal{I}$ classically satisfies~$F$, i.e., $\mathrm{deg}_\mathcal{I}(F) < \infty$, we get $\mathrm{deg}_\mathcal{I}(F \x G) = \mathrm{deg}_\mathcal{I}(F) \leq \mathrm{opt}(F)$; if $\mathcal{I}$ does not classically satisfy~$F$, i.e., $\mathrm{deg}_\mathcal{I}(F) = \infty$, we get $\mathrm{deg}_\mathcal{I}(F \x G) = \mathrm{opt}(F) + \mathrm{deg}_\mathcal{I}(G) > \mathrm{opt}(F)$. 

We now define the central notion of preferred models, and then give a small example of QCL-semantics in action.

\begin{definition}
    Let $F$ be a QCL-formula. $\mathcal{I}$ is a preferred model of $F$ iff $\mathrm{deg}_\mathcal{I}(F) < \infty$ and $\mathrm{deg}_\mathcal{I}(F) \leq \mathrm{deg}_\mathcal{J}(F)$ for all other interpretations $\mathcal{J}$.
\end{definition}

\begin{example}
    The QCL-formula $F = (a \land b) \x a \x b$ expresses that satisfying both $a$ and $b$ is preferable to satisfying only $a$, which in turn is preferable to satisfying only $b$. First, observe that $\mathrm{opt}(F) = 3$. Moreover, $\mathrm{deg}_\emptyset(F) = \infty$, $\mathrm{deg}_{\{b\}}(F) = 3$, $\mathrm{deg}_{\{a\}}(F) = 2$, and $\mathrm{deg}_{\{a,b\}}(F) = 1$. Thus, $\{a,b\}$ is a preferred model of $F$. 
    
    Now consider $F' = ((a \land b) \x a \x b) \land \neg (a \land b)$, which is similar to $F$, but with the additional information that $a$ and $b$ can not be jointly satisfied. Again, $\mathrm{deg}_\emptyset(F') = \infty$, $\mathrm{deg}_{\{b\}}(F') = 3$, and $\mathrm{deg}_{\{a\}}(F') = 2$. However, $\mathrm{deg}_{\{a,b\}}(F') = \infty$, i.e., $\{a,b\}$ does not satisfy $F'$. Since it is not possible to satisfy $F'$ to a degree of $1$, $\{a\}$ is a preferred model of $F'$.
\end{example}

Note that ordered disjunction is associative under QCL-semantics, which means that we can simply write $A_1 \x A_2 \x \ldots \x A_n$ to express that we must satisfy at least one of $A_1,\ldots,A_n$, and that we prefer $A_i$ to $A_j$ for $i < j$. 
Formally, this is expressed by the following lemma:
\begin{lemma}[from \cite{BrewkaEtAl2004qcl}]\label{ordered-disjunction-associative}
    Let $F$, $G$, and $H$ be QCL-formulas. Then $(F \x (G \x H))$ and $((F \x G) \x H)$ have the same optionality and the same satisfaction degree under all interpretations. 
\end{lemma}

As mentioned in the introduction, an alternative semantics for QCL has been proposed in the form of PQCL \cite{BenferhatSedki2008pqcl}. Specifically, PQCL changes the semantics for the classical connectives ($\neg,\lor,\land$), but defines ordered disjunction ($\x$) in the same way as QCL. For our purposes, it is not necessary to formally define PQCL. Rather, it suffices to note that, in PQCL, negation propagates to the atom level, meaning that $\neg (F \land G)$ is simply assigned the satisfaction degree of $\neg F \lor \neg G$, $\neg (F \lor G)$ is assigned the degree of $\neg F \land \neg G$, and $\neg (F \x G)$ is assigned the degree of $\neg F \x \neg G$. 

\subsection{Game-Theoretic Semantics (GTS)}
We start by recalling Hintikka's game \cite{Hintikka1973-HINLLA} over a formula $F$ in the language restricted to the connectives $\vee, \wedge, \neg$ and over an interpretation $\mathcal{I}$. The game is played between two players, \Me and \You, both of which can act either in the role of Proponent ($\mathbf{P}$) or Opponent ($\mathbf{O}$). The game starts with \Me as $\mathbf{P}$ of the formula $F$ and \You as $\mathbf{O}$. At formulas of the form $G_1 \vee G_2$, $\mathbf{P}$ chooses a formula $G_i$ that the game  continues with. At formulas of the form $G_1 \wedge G_2$ it is $\mathbf{O}$'s choice. At negations $\neg G$, the game continues with $G$ and a role switch. Every outcome of the game is a propositional variable $a$. The player currently in the role of $\mathbf{P}$ wins the game (and $\mathbf{O}$ loses) iff $a\in \mathcal{I}$. It is known that \I have a winning strategy for this game iff $\mathcal{I}\models F$.

How can we extend Hintikka's game from classical logic to choice logic? We propose the following intuitive reading of ordered disjunction ($\x$):
at $G_1 \x G_2$ it is $\mathbf{P}$'s choice whether to continue with $G_1$ or with $G_2$, \textit{but this player prefers} $G_1$.
\My aim in the game is now not only to win the game but to do so with as little compromise to \My preferences as possible. Thus, it is natural to express \My preference of $G_2$-outcomes $O_2$ over $G_1$-outcomes $O_1$ via the relation $O_1 \ll O_2$. We leave the formal treatment of this game for the next section and conclude with some standard game-theoretic definitions.

\begin{definition}
A \emph{game} is a pair $\mathbf{G} = (T,d)$, where
\begin{enumerate}
\item $T=(V,E,l)$ is a tree with set of nodes $V$ (usually called \emph{(game) states}) and edges $E$. The leafs of $T$ are called \emph{outcomes} and are denoted $\mathcal{O}(\mathbf{G})$. The \emph{labelling function} $l$ maps nodes of $T$ to the set $\{I,Y\}$.
\item $d$ is a payoff-function mapping outcomes to elements of a linear order $(D,\preceq)$. 
\end{enumerate}
We write $x\approx y$ if $x\preceq y$ and $y\preceq x$.  In the games that we are interested in, $D$ is partitioned into two sets, $W$ and $L$, where $W$ is upward-closed and $L = {D \setminus W}$. Outcomes $O$ are called \emph{winning} if $d(O) \in W$ and \emph{losing} if $d(O)\in L$. 
\end{definition}

For example, in Hintikka's game (without negations), the underlying game tree is exactly the tree-representation of the formula $F$. The labels of $\wedge$-formulas are ``Y'', while the label of $\vee$-formulas are ``I''. The payoff functions maps outcomes to $P=\{0,1\}$, where $d(p) = 1$ iff $p \in \mathcal{I}$. $P$ carries the usual ordering $0<1$ and $W = \{1\}$.

A strategy $\sigma$ for \Me in a game can be understood as \My complete game-plan. For every node of the underlying game-tree labelled ``I'', $\sigma$ tells \Me to which node \I have to move. Here is a formal definition:

\begin{definition}
A \emph{strategy} $\sigma$ for \Me for the game $\mathbf{G}$ is a subset of the nodes of the underlying tree such that (1) the root of $T$ is in $\sigma$ and for all $v\in \sigma$, (2) if $l(v) = I$, then at least one successor of $v$ is in $\sigma$ and (3) if $l(v) = Y$, then all successors of $v$ are in $\sigma$. A strategy for \You is defined symmetrically. We denote by $\Sigma_I$ and $\Sigma_{Y}$ the set of all strategies for \Me and \You, respectively.
\end{definition}

Conditions (1) and (3) make sure that all possible moves by the other player are taken care of by the game-plan.

Note that each pair of strategies $\sigma_I \in \Sigma_{I}$, $\sigma_{Y} \in \Sigma_{Y}$ defines a \emph{unique} outcome of $\mathbf{G}$, which we will denote by $O(\sigma_I,\sigma_{Y})$. We abbreviate $d(O(\sigma_I,\sigma_{Y}))$ by $d(\sigma_I,\sigma_{Y})$. A strategy $\sigma_I^*$ for \Me is called \emph{winning} if, playing according to this strategy, \I win the game, no matter how \You move, i.e. for all $\sigma_{Y} \in \Sigma_{Y}$, $d(\sigma_I^*, \sigma_{Y})\in W$. An  outcome $O$ that maximizes \My pay-off in light of \Your best strategy is called \emph{maxmin-outcome}. Formally, $O$ is a maxmin-outcome iff $d(O) \approx \max^{\preceq}_{\sigma_I} \min^\preceq_{\sigma_{Y}} d(\sigma_I,\sigma_{Y})$ and $d(O)$ is called the \emph{maxmin-value} of the game. A strategy $\sigma_I^*$ for \Me is a \emph{maxmin-strategy} for $\mathbf{G}$ if $\sigma_I^* \in \mathrm{arg}\max^{\preceq}_{\sigma_I} \min^\preceq_{\sigma_{Y}} d(\sigma_I,\sigma_{Y})$. Minmax values and strategies for \You are defined symmetrically. 

The class of games that we have defined falls into the category of \emph{zero-sum games} in game-theory. They are characterized by the fact that the players have strictly opposing interests. In zero-sum games, the minmax and maxmin value always coincide and is referred to as the \emph{value of the game}.

\section{A Game for QCL} \label{sec:capturing-qcl}
We now give a formal definition of a Hintikka-style game for QCL. As motivated in the previous section, we seek to capture the intuition that ordered disjunction ($\x$) should be interpreted as a preference of a player for all outcomes on the left of $\x$ over all outcomes on the right. Game states will be of the form $\mathbf{P}:F$ or $\mathbf{O}:F$, where $F$ is a QCL-formula and the labels ``$\mathbf{P}$'' and ``$\mathbf{O}$'' are to signify that \I currently act in the role of proponent and opponent, respectively. 

We inductively define the game tree $T(\mathbf{Q}:F)$ of the game $\mathbf{G}(\mathbf{Q}:F,\mathcal{I})$ for $\mathbf{Q}\in \{\mathbf{P},\mathbf{O}\}$, as well as an order $\ll$ on outcomes, which represents preferences in the game from \My viewpoint (\Yours are the exact opposite). Both definitions will be independent of the given interpretation $\mathcal{I}$. After that we define the payoff-function $d$, which respects $\ll$ on \My winning outcomes. The definition of $T(\mathbf{Q}:F)$ depends on the structure of $F$:

\begin{description}
\item[$(R_a)$] ${T}(\mathbf{P}:a)$ consists of the single leaf $r$ and $\ll_{\mathbf{P}:a} = \emptyset$.
\item[$(R_\neg)$] ${T}(\mathbf{P}:\neg G)$, consists of a root $r$ labelled ``I'', and immediate subtree ${T}(\mathbf{O}:G)$ and $\ll_{\mathbf{P}:\neg G} = \emptyset$, i.e. at $\mathbf{P}:\neg G$, the game continues with a role switch and erased preferences for the remainder of the game.
\item[$(R_\wedge)$] ${T}(\mathbf{P}: G_1 \wedge G_2)$ is a tree with root $r$ labelled ``Y'', and immediate subtrees ${T}(\mathbf{P}:G_1)$ and ${T}(\mathbf{P}:G_2)  $, i.e. at $\mathbf{P}:G_1 \wedge G_2$, \You choose whether to continue with $\mathbf{P}:G_1$ or with $\mathbf{P}:G_2$. The preference is given by $\ll_{\mathbf{P}:G_1 \wedge G_2} = \ll_{\mathbf{P}:G_1 } \cup \ll_{\mathbf{P}:G_2}$.
\item[$(R_\vee)$] ${T}(\mathbf{P}: G_1 \vee G_2)$ is a tree with root $r$ labelled ``I'', and immediate subtrees ${T}(\mathbf{P}:G_1)$ and ${T}(\mathbf{P}:G_2)  $, i.e. at $\mathbf{P}:G_1 \vee G_2$, \I choose whether to continue with $\mathbf{P}:G_1$ or with $\mathbf{P}:G_2$. The preference is given by $\ll_{\mathbf{P}:G_1 \wedge G_2} = \ll_{\mathbf{P}:G_1 } \cup \ll_{\mathbf{P}:G_2}$.
\item[$(R_{\x})$] ${T}(\mathbf{P}: G_1 \x G_2)$ is a tree with root $r$ labelled ``I'', and immediate subtrees ${T}(\mathbf{P}:G_1)$ and ${T}(\mathbf{P}:G_2)  $, i.e. at $\mathbf{P}:G_1 \x G_2$, \I choose whether to continue with $\mathbf{P}:G_1$ or with $\mathbf{P}:G_2$. The preference is given by $O_1 \ll_{\mathbf{P}:G_1 \x G_2} O_2$ iff $O_1 \in \mathcal{O}(\mathbf{P}:G_2)$\footnote{For simplicity, we write $\mathcal{O}(\mathbf{Q}:F)$ instead of $\mathcal{O}(\mathbf{G}(\mathbf{Q}:F))$} and $O_2 \in \mathcal{O}(\mathbf{P}:G_1)$, or $O_1 \ll_{\mathbf{P}:G_j} O_2$ for $j\in\{1,2\}$. This means that \I prefer all winning outcomes of the $G_1$-game over all outcomes of the $G_2$-game.
\end{description}

The tree $T(\mathbf{O}:F)$ is the same as ${T}(\mathbf{P}:F)$, except that labels are swapped and the preference relation is always empty. For example, $T(\mathbf{O}:G_1\x G_2)$ consists of the node labelled ``Y'' with immediate subtrees $T(\mathbf{O}:G_1)$ and $T(\mathbf{O}:G_2)$ and $\ll_{\mathbf{O}:G_1\x G_2} = \emptyset$. The rule $(R_\neg)$ may sound a bit counter-intuitive, but it precisely captures negation in QCL. For a critical discussion on negation, see Subsetion~\ref{subsect:issues}.

Let us say that an \emph{atomic} game state $\mathbf{P}: a$  is true in $\mathcal{I}$ if $a \in \mathcal{I}$, and false otherwise. Conversely, $\mathbf{O}: a$ is true if $a \not\in \mathcal{I}$, and false otherwise. The payoff-function $d$ is defined as follows: given an outcome $O$, let $\pi_\ll(O) = \{O_1,O_2, ..., O_n\}$ be the longest $\ll$-path starting in $O$, i.e. $O = O_1$, the $O_i$ are pairwise different outcomes and $O_i \ll O_{i+1}$ for all $1 \leq i \leq n-1$. Let us say that is  $O$ is true, if $O$ stands for a true atomic game state, and otherwise false. The payoff-function $d_\mathcal{I}$ maps into the set $D = \mathbb{N}\cup \{\infty\}$, linearly ordered by $\preceq$, the inverse natural ordering (1 is best, $\infty$ is worst):
\[d_\mathcal{I}(O) = \begin{cases}
				|\pi_\ll(O)|, &\text{ if } O \text{ is true},\\
				\infty, &\text{ if } O \text{ is false}.\\
\end{cases}\]If $\mathcal{I}$ is clear from context, we simply write $d$ instead of $d_\mathcal{I}$.

\begin{figure*}[t]
    \begin{center}
    {\begin{tikzpicture}[level distance = 0.9cm]
    \tikzstyle{level 1}=[sibling distance=55mm]
    \tikzstyle{level 2}=[sibling distance=40mm]
    \tikzstyle{level 2}=[sibling distance=25mm]
    \tikzstyle{level 6}=[sibling distance=30mm]
        \node {\(\left[\mathbf{P}:((a \x b) \x c) \land \neg (a \x d) \right]^{Y}\)}
        	child {node {\(\left[\mathbf{P}:(a \x b) \x c \right]^I\)}
                child {node {\(\left[\mathbf{P}:a \x b \right]^I\)}
                    child {node {\(\left[\mathbf{P}:a \right]\)}}
                    child {node {\(\left[\mathbf{P}:b \right]\)}}
                }
                child {node {\(\left[\mathbf{P}:c \right]\)}}
            }
            child {node {\(\left[\mathbf{P}:\neg (a \x d) \right]^I\)}
                child {node {\(\left[\mathbf{O}:a \x d \right]^{Y}\)}
            		child {node {\(\left[\mathbf{O}:a \right]\)}}
                    child {node {\(\left[\mathbf{O}:d \right]\)}}
                }
            };
    		\end{tikzpicture}}
    \end{center}
    \caption{A game tree for QCL.} 
    \label{fig:treeFirstNegation}
\end{figure*}
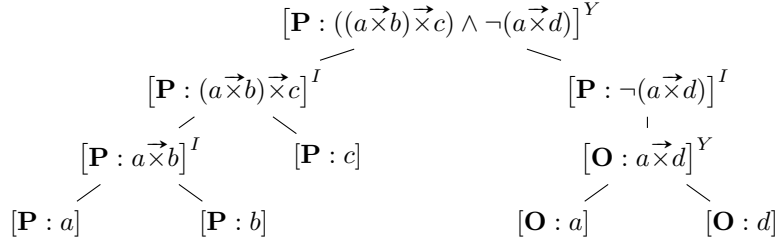

\begin{example}\label{ex:gameFirstNegation}
    Consider the formula $F = ((a \x b) \x c) \land \neg (a \x d)$. Figure~\ref{fig:treeFirstNegation} depicts the corresponding game tree. Observe that the node ${\mathbf{O}: a \x d}$  has the label ``Y'' because the roles of the players are switched in the parent node ${\mathbf{P}:\neg (a \x d)}$. The order on outcomes is ${\mathbf{P}:c} \ll {\mathbf{P}:b} \ll {\mathbf{P}:a}$. Note that ${\mathbf{O}:d} \centernot\ll {\mathbf{O}:a}$, since preferences are deleted via the negation rule $(R_\neg)$.
    
    Consider the interpretation $\{a\}$. The winning outcomes are ${\mathbf{P}:a}$ and ${\mathbf{O}:d}$ and the degree-order is given by $d({\mathbf{P}:a}) = 1$, $d({\mathbf{P}:b}) = d({\mathbf{P}:c}) = d({\mathbf{O}:a}) = \infty$, $d({\mathbf{O}:d}) = 2$.  
    In this case, \I have no winning strategy: in the root node, it is \Your turn, and \You can move to the right where the leaf ${\mathbf{O}:a}$ is not a winning outcome because roles were switched.
    
   Now consider $\{b\}$. The winning outcomes are ${\mathbf{P}:b}$, ${\mathbf{O}:a}$ and ${\mathbf{O}:d}$ and the degree-order is given by $d({\mathbf{P}:a}) = d({\mathbf{P}:c}) = \infty$, $d({\mathbf{P}:b}) = 2$, $d({\mathbf{O}:a}) = d({\mathbf{O}:d}) = 1$. If \You move to the right this only leads to winning outcomes (with value $1$) for \Me (due to switched roles). If \You move to the left, it is now always \My turn, and \I can reach the winning outcome ${\mathbf{P}:b}$ with a value of $2$. Thus, it is better for \You to move to the left, and the best outcome for \Me is ${\mathbf{P}:b}$.
\end{example}

\begin{lemma}\label{longestpathopt}
The longest $\ll$-path in $\mathcal{O}(\mathbf{P}:F)$ has length $\mathrm{opt}(F)$.
\end{lemma}


The above Lemma can be shown via induction on $F$, using the fact that preferences are erased at negations as well as in all game trees $T(\mathbf{O}:G)$. 
This is already quite a nice result as it shows that the notion of optionally arises naturally in our game, whereas in QCL, optionality  must be defined a-priori to ensure that the semantics work as intended.
We are now ready to show that our game semantics captures QCL:

\begin{theorem}\label{thm:adeqqcl}
The value of $\mathbf{G}(\mathbf{P}:F,\mathcal{I})$ is equal to $\dI(F)$.
\end{theorem}
 
For the proof of Theorem~\ref{thm:adeqqcl}, we introduce some handy notation:  When $\mathcal{I}$ is clear from context, we denote by $O({\mathbf{Q}:F})$ the maxmin-outcome and by $d({\mathbf{Q}:F})$ the maxmin-value of the game $\mathbf{G}({\mathbf{Q}:F},\mathcal{I})$. Where it does not cause confusion, we will identify a formula with the corresponding node in the game tree. Since the payoff-function $d$ differs from one game to another, let us denote by $d_{{\mathbf{Q}:F}}$ the payoff-function for the game $\mathbf{G}({\mathbf{Q}:F},\mathcal{I})$.
 
\begin{proof}[Proof (of Theorem~\ref{thm:adeqqcl})]
It suffices to show the following two claims by induction on $F$: (1) $d(\mathbf{P}:F)=\dI(F)$ and (2) $d(\mathbf{O}:F) = \infty$, if $\dI(F)<\infty$ and $1$ otherwise. Since in the game $\mathbf{G}(\mathbf{O}:F,\mathcal{I})$ all preferences are deleted, it is essentially Hintikka's game. We will therefore prove only (1). Remember that the ordering $\preceq$  is the inverse of the natural ordering $\leq$ on $\mathbb{N}\cup\{\infty\}$. \I therefore seek to $\leq$-\emph{minimize} \My payoff in the game.

$F = a$: This game consists of a single node $v$. The longest $\ll$-path starting at $v$ has length 1. Therefore, $d({\mathbf{P}:a}) = 1$ iff $a\in \mathcal{I}$ iff $\dI(a) = 1$ and $d({\mathbf{P}:a}) = \infty$ iff $a\notin \mathcal{I}$ iff $\dI(a) = \infty$.

$F = G_1 \wedge G_2$: In the first round, \You choose between $\mathbf{P}:G_1$ and $\mathbf{P}: G_2$. \Your best strategy is to go to the subgame with $\preceq$-minimal payoff:
\begin{align*}
d({\mathbf{P}:G_1 \wedge G_2})
&= \min_{\preceq}\{d({\mathbf{P}:G_1}),d({\mathbf{P}:G_2})\}\\
&= \max \{\dI(G_1),\dI(G_2)\} \\
&= \dI(G_1\wedge G_2)
\end{align*}Here, the second step used the fact that $\preceq$ is the inverse of the natural ordering on $\mathbb{N}\cup\{\infty\}$ and 
the induction hypothesis.

$F = G_1 \vee G_2$: In the first round, \I choose between $\mathbf{P}:G_1$ and $\mathbf{P}:G_2$, Therefore:
\begin{align*}
d(\mathbf{P}:G_1 \vee G_2)
&= \max_{\preceq}\{d(\mathbf{P}:G_1),d({\mathbf{P}:G_2})\}\\
&= \min \{\dI(G_1),\dI(G_2)\} \\
&= \dI(G_1\wedge G_2)
\end{align*}
Again, we used the induction hypothesis in the second step.

$F = G_1 \x G_2$: In the first round, \I choose between $\mathbf{P}:G_1$ and $\mathbf{P}:G_2$, but all outcomes of the $G_2$-game are in $\ll$-relation to all outcomes of the $G_1$-game, which is respected by the payoff-function for the winning outcomes. Let us write $W(G_i)$ for the winning outcomes of $\mathbf{G}(\mathbf{P}:G_i,\mathcal{I})$, i.e. those outcomes $O$ with $d_{\mathbf{P}:G_i}(O) < \infty$. By Lemma~\ref{longestpathopt}, the longest $\ll$- path in $\mathcal{O}(\mathbf{P}:G_1)$ has length $\mathrm{opt}(G_1)$. Hence, for all $O\in \mathcal{O}(\mathbf{P}:G_1 \x G_2)$:
\[d_{\mathbf{P}:F}(O) = 
\begin{cases}
    d_{\mathbf{P}:G_1}(O), &\text{ if } O\in  W(G_1),\\
    d_{\mathbf{P}:G_2}(O)+\mathrm{opt}(G_1), &\text{ if } O\in  W(G_2),\\
    \infty, &\text{ otherwise}.
\end{cases}\]
Therefore, if $O(\mathbf{P}: G_1)$ is winning, i.e., $d({\mathbf{P}:G_1})=\dI(G_1)<\infty$, then \I move to $\mathbf{P}:G_1$ and $d({\mathbf{P}:G_1 \x G_2}) = d_{\mathbf{P}:F}(O(\mathbf{P}:F)) = d_{\mathbf{P}:G_1}(O(\mathbf{P}: G_1))=\dI(G_1) = \dI(F)$. If $O(\mathbf{P}:G_1)$ is losing, but $O(\mathbf{P}:G_2)$ is winning, i.e., $d({\mathbf{P}:G_1})=\dI(G_1) = \infty$ and $d({\mathbf{P}:G_2}) = \dI(G_2) < \infty$, then \I move to $\mathbf{P}:G_2$ and $d({\mathbf{P}:F})=d_{\mathbf{P}:F}(O(\mathbf{P}:F)) = d_{\mathbf{P}: G_2}(O(\mathbf{P}: G_2)) + \mathrm{opt}(G_1) = \dI(G_2) + \mathrm{opt}(G_1) = \dI(F)$. If both $O(\mathbf{P}: G_1)$ and $O(\mathbf{P}: G_2)$ are losing, then $d({\mathbf{P}:F})= \infty = \dI(F)$.

$F = \neg G$: Here the game continues with $\mathbf{O}:G$. By (2), $d({\mathbf{O}:G}) = \infty$, if $\dI(G)<\infty$ and $1$ otherwise. Hence, $d({\mathbf{P}:\neg G}) = d({\mathbf{O}:G}) = \dI(\neg G)$.
\end{proof}

\section{A New Semantics}

In this section, we first identify some contentious behavior in QCL (and PQCL) with regards to negation. We then address these issues by adapting our game semantics from Section~\ref{sec:capturing-qcl} to use GTS negation. Lastly, we specify a degree-semantics for this new game.

\subsection{Negation in QCL (and PQCL)} \label{subsect:issues}

While choice logics are a useful formalism to express both soft constraints (preferences) and hard constraints (truth) in a single language, existing semantics (such as QCL and PQCL) are not without problems. One matter of contention lies in how choice logics deal with negation, as we will now see. 

Consider the following statement: \textit{``I want almond or banana ice cream, but preferably almond''}. It is clear how to formalize this sentence in QCL, namely simply as $a \x b$, where $a$ and $b$ are abbreviations for almond and banana ice cream respectively. However, it is not obvious, and generally not agreed upon, how the negation of this sentence is to be understood. In QCL, negation erases all information about preferences and only takes classical satisfaction into account. Under this interpretation, $\neg (a \x b)$ is to be read as \textit{``I do not want almond or banana ice cream''}, which means that $\neg (a \x b)$ is equivalent to $\neg a \land \neg b$ (cf.\ Table~\ref{table:negation-comparison}). This is certainly a pragmatic approach, but, as we believe, potentially problematic because it means that negation does not actually apply to the \emph{entire} sentence, i.e., $\neg(a \x b)$ can not be read as \textit{``I do not want almond or banana ice cream, but preferably almond''}. Moreover, this approach means that a QCL-formula $F$ is not logically equivalent to its double negation $\neg \neg F$, which may lead to non-intuitive behavior. Consider the following sentence: \textit{``I do not not want almond or banana icecream, and preferably almond''}. It is certainly reasonable to understand this sentence to be equivalent to
our initial sentence.
However, $\neg \neg (a \x b)$ is not logically equivalent to $(a \x b)$ in QCL since all information about preferences is lost in $\neg \neg (a \x b)$. 

PQCL addresses this issue, i.e., $F$ is always equivalent to $\neg\neg F$. However, in our view, PQCL introduces two other problems with negation. In PQCL, $\neg (a \x b)$ is interpreted as $\neg a \x \neg b$ (cf.\ Table~\ref{table:negation-comparison}). 
The first issue is that it is possible for an interpretation to classically satisfy both a formula $F$ and its negation $\neg F$. For example, in Table~\ref{table:negation-comparison} we can see that the interpretations $\{a\}$ and $\{b\}$ classically satisfy both $(a \x b)$ and $\neg (a \x b)$. However, the ordered disjunct $(a \x b)$ not only expresses a preference, but also a hard constraint ($a$ or $b$ must be satisfied). We believe that negation should act on both of these aspects. Moreover, as a result of this behavior, the implication $F \rightarrow G$ can in general not be defined via $\neg F \lor G$. For example, the formula $\neg (a \x b) \lor c$ is classically satisfied by the interpretations $\{a\}$ and $\{b\}$, although the antecedent $(a \x b)$ is satisfied under these interpretations while the consequent $c$ is not.
Secondly, the satisfaction degree of $\neg F$ does not only depend on the satisfaction degree and optionality of $F$.  Looking again at Table ~\ref{table:negation-comparison} we see that the interpretations $\{a\}$ and $\{a,b\}$ satisfy $(a \x b)$ to the same degree, i.e., they are equally preferable, but $\{a\}$ satisfies $\neg (a \x b)$ to a degree of $2$ while $\{a,b\}$ does not satisfy $\neg (a \x b)$ at all.

\begin{table}[t]
    \centering
    \begin{tabular}{c|ccc} 
        $\mathcal{I}$ & $a \x b$ & $\neg a \land \neg b$ & $\neg a \x \neg b$ \\ 
        \hline
        $\emptyset$ & $\infty$ & $1$ & $1$ \\ 
        $\{b\}$ & $2$ & $\infty$ & $1$ \\
        $\{a\}$ & $1$ & $\infty$ & $2$ \\
        $\{a,b\}$ & $1$ & $\infty$ & $\infty$
    \end{tabular}
    \caption{Truth table showing the satisfaction degrees of $\neg (a \protect\x b)$ in QCL (equivalent to $\neg a \land \neg b$) and PQCL (equivalent to $\neg a \protect\x \neg b$).}
    \label{table:negation-comparison}
\end{table}

When designing our new game semantics, we will keep these issues in mind. Our main goal is to define a negation that acts both on hard-constraints (truth) as in QCL and soft-constraints (preferences) as in PQCL. Moreover, we will ensure that (1) formulas are equivalent to their double negation, (2) formulas and their negation can not be satisfied classically by the same interpretation, and (3) the satisfaction degree of $\neg F$ depends only on the satisfaction degree of $F$. 

It must be noted that QCL and PQCL are not the only semantics for propositional logic extended with ordered disjunction. Another example can be found in \cite{MalyWoltran2018hardsoft}, where the concept of satisfaction degrees is abandoned. The semantics induce a partial order among interpretations. 
However, negation is handled in the same way as in QCL, i.e., all information about preferences is lost and therefore formulas are not equivalent to their double negation.

\subsection{Using GTS Negation} \label{subsec:new-game}

We now address the issues with negation in previous approaches to choice logics, as outlined in Section~\ref{subsect:issues}. One of the main problems was that all information on preferences is lost in negated formulas. From the game-point of view, this corresponds to the fact that at game states $\neg F$, all preferences are deleted. But in a game-theoretic setting it is natural to consider a game where this deletion does not occur. We therefore propose the following rule for negation:

\begin{description}
\item[$(R_\neg)$] ${T}(\mathbf{P}:\neg G)$, consists of a root $r$ labelled ``I'', the immediate subtree ${T}(\mathbf{O}:G)$, and $\ll_{\mathbf{P}:\neg G}$ equal to $\ll_{\mathbf{O}:G}$, i.e., at $\mathbf{P}:\neg G$, the game continues with a role switch.
\end{description}
 
All other rules stay the same, except that in $T(\mathbf{O}:F)$ labels are swapped and preferences are switched, i.e. $\ll_{\mathbf{O}:F}$ is the inverse of $\ll_{\mathbf{P}:F}$. For example, if $F = G_1 \x G_2$, then $T(\mathbf{O}:F)$ consists of the node labelled ``Y'' with immediate subtrees $T(\mathbf{O}:G_1)$ and $T(\mathbf{O}:G_2)$ and $O_1 \ll_{\mathbf{O:F}} O_2$ iff $O_2 \ll_{\mathbf{P:F}} O_1$.

Additionally, we change our payoff-function to respect preferences not only in Player I's winning outcomes, but in both Player's winning outcomes:

\begin{definition}\label{ex:newdeg}
Let $Z := (\mathbb{Z} \setminus \{0\},\trianglelefteq)$. The ordering $\trianglelefteq$ is the inverse of the natural ordering on $\mathbb{Z}^-$ and on $\mathbb{Z}^+$ and for $a \in \mathbb{Z}^+, b\in \mathbb{Z}^-$ we set $b \triangleleft a$. For an outcome $O$, we set\footnote{Notice the flipped $\ll$-sign in the second case.}
 \[\delta_\mathcal{I}(O) = \begin{cases}
				|\pi_\ll(O)|, &\text{ if } O \text{ is true},\\
				-|\pi_\gg(O)|, &\text{ if } O \text{ is false}.\\
\end{cases}\]
Again, we write $\delta$, instead of $\delta_\mathcal{I}$, if $\mathcal{I}$ is clear from context.
We denote the new game initiated at the game state $\mathbf{Q}:F$ and played over the interpretation $\mathcal{I}$ by $\mathbf{NG}(\mathbf{Q}:F,\mathcal{I})$.
\end{definition}

\begin{figure*}[t]
\includegraphics[width = \textwidth]{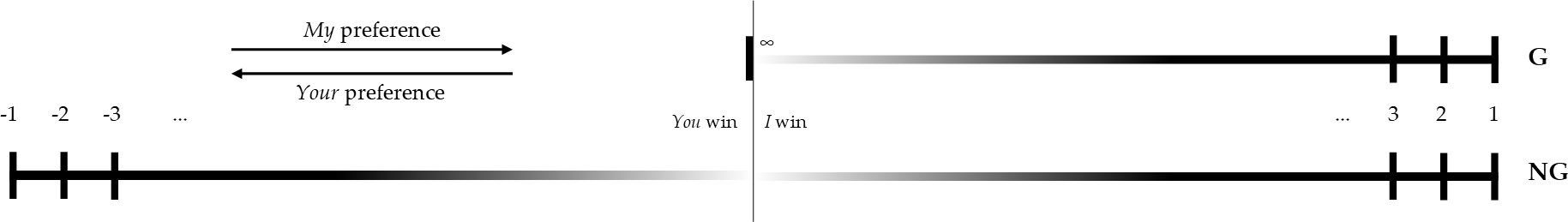}
\caption{Preferences and winning payoffs of the two players in the games $\mathbf{G}$ and $\mathbf{NG}$}
\label{fig:newwinningrange}
\end{figure*}

Observe that with these alterations, the two players become truly symmetric. The goal of \emph{both} players is now to (1)~win the game with (2)~as little compromise as possible, and otherwise (3)~force the opponent in as much compromise as possible. See Figure~\ref{fig:newwinningrange} for an instructive graphical representation, which also shows how this new approach differs from the old approach of QCL and its corresponding game~$\mathbf{G}$.

\begin{example}\label{ex:gameSecondNegation}
Consider again the the formula $((a \x b) \x c) \land \neg (a \x d)$ from Example~\ref{ex:gameFirstNegation}. The game tree is the same as before (see Figure~\ref{fig:treeFirstNegation}). The order on outcomes is now ${\mathbf{P}:c} \ll {\mathbf{P}:b} \ll {\mathbf{P}:a}$ and ${\mathbf{O}:a} \ll {\mathbf{O}:d}$.  
    
    Consider the interpretation $\{a\}$. As in Example~\ref{ex:gameFirstNegation}, \I have no winning strategy since \You can always move to the right side at the root node and to the left side in ${\mathbf{O}:a \x d}$ to reach ${\mathbf{O}:a}$ with a value of $-1$.
    
    Now consider $\{d\}$. The only winning outcome is ${\mathbf{O}:a}$ and the payoffs are given by $\delta({\mathbf{P}:c}) = -1$, $\delta({\mathbf{P}:b})  = -2$, $\delta({\mathbf{P}:a}) = -3$, $\delta({\mathbf{O}:a}) = 2$, $\delta({\mathbf{O}:d}) = -2$. If \You move to the left at the root node, it is best for \Me to reach the outcome ${\mathbf{P}:a}$ with a payoff of $-3$. If \You move to the right at the root node, \You have to move to the right again in ${\mathbf{O}:a \x d}$ to reach a losing outcome, namely ${\mathbf{O}:d}$ with a payoff of $-2$. Thus, it is better for \You to move to the right at the root note, giving us the game value $-2$. 
\end{example}

\subsection{Extracting a Degree Semantics} \label{subsec:extracting-degree-semantics}

Using our game $\mathbf{NG}$ as a cornerstone, we now define a degree-function for QCL-formulas taking values in the domain $(Z  ,\trianglelefteq)$ from Definition~\ref{ex:newdeg} and discuss some of its properties. The resulting logic will be called GCL, for Game-induced Choice Logic. The proof of adequacy of the game $\mathbf{NG}$ with respect to this degree-function is topic of the next subsection.
We denote the optionality function of GCL by $\mathrm{opt}^G$, and define it in the same way as the optionality function $\mathrm{opt}$ of QCL, except for negation:
\[\mathrm{opt}^G(\neg F) = \mathrm{opt}^G(F)\]
With this definition, we can extend Lemma~\ref{longestpathopt} for $\mathbf{NG}$:

\begin{lemma}\label{lem:longestpathng}
    Let $\mathbf{Q} \in \{\mathbf{P},\mathbf{Q}\}$. The longest $\ll$-path in $\mathcal{O}({\mathbf{Q}:F})$ has length $\mathrm{opt}^G(F)$.
\end{lemma}

The degree function of GCL is denoted by $\mathrm{deg}_\mathcal{I}^G$. It assigns to each formula a degree relative to an interpretation $\mathcal{I}$ and is defined inductively as follows (for succinctness, we abbreviate $\mathrm{opt}^G(F)$ with $o_F$ and $\mathrm{opt}^G(G)$ with $o_G$):
\begin{align*}
    \mathrm{deg}_\mathcal{I}^G(a) &= 
        1 \text{ if } a \in \mathcal{I}, -1 \text{ otherwise } \\
    \mathrm{deg}_\mathcal{I}^G(\neg F) &= - \mathrm{deg}_\mathcal{I}^G(F)\\
    \mathrm{deg}_\mathcal{I}^G(F\wedge G) &= \min(\mathrm{deg}_\mathcal{I}^G(F),\mathrm{deg}_\mathcal{I}^G(G))\\
    \mathrm{deg}_\mathcal{I}^G(F\vee G) &= \max(\mathrm{deg}_\mathcal{I}^G(F),\mathrm{deg}_\mathcal{I}^G(G)) \\
    \mathrm{deg}_\mathcal{I}^G(F\x G) &= 
        \begin{cases}
			\mathrm{deg}_\mathcal{I}^G(F) & \text{if } \mathrm{deg}_\mathcal{I}^G(F) \in \mathbb{Z}^+\\
            o_F + \mathrm{deg}_\mathcal{I}^G(G) & \text{if } \mathrm{deg}_\mathcal{I}^G(F) \in \mathbb{Z}^-, \\ & \mathrm{deg}_\mathcal{I}^G(G) \in \mathbb{Z}^+\\
            \mathrm{deg}_\mathcal{I}^G(F)-o_G & \text{otherwise}
                                    \\\end{cases}
\end{align*}
Here $\min$ and $\max$ are relative to $\trianglelefteq$. If $\mathrm{deg}_\mathcal{I}^G(F) \in \mathbb{Z}^+$ then we say that $\mathcal{I}$ classically satisfies $F$, or that $\mathcal{I}$ is a model of $F$. Note that, in contrast to QCL, those interpretations that result in a higher degree relative to the ordering $\trianglelefteq$ are more preferable, which is also why we take the maximum degree for disjunction and the minimum degree for conjunction. However, since $\trianglelefteq$ inverts the order on $\mathbb{Z}^+$, a degree of $1$ is considered to be higher than a degree of $2$. With this in mind, the notion of preferred models can be defined analogously to QCL:

\begin{definition}
    Let $F$ be a QCL-formula. Under our new semantics, $\mathcal{I}$ is a preferred model of $F$ iff $\mathrm{deg}_\mathcal{I}^G(F) \in \mathbb{Z}^+$ and $\mathrm{deg}_\mathcal{J}^G(F) \trianglelefteq \mathrm{deg}_\mathcal{I}^G(F)$ for all other interpretations $\mathcal{J}$.
\end{definition}

First, we show that ordered disjunction is still associative under these new semantics:

\begin{lemma}\label{lemma:new-semantics-associativity}
    Let $F_1 = ((A \x B) \x C)$, $F_2 = (A \x (B \x C))$ for any QCL-formulas $A,B,C$. Then $\mathrm{opt}^G(F_1) = \mathrm{opt}^G(F_2)$ and $\mathrm{deg}_\mathcal{I}^G(F_1) = \mathrm{deg}_\mathcal{I}^G(F_2)$ for all interpretations~$\mathcal{I}$.
\end{lemma}
\begin{proof}
    $\mathrm{opt}^G(F_1) = \mathrm{opt}^G(F_2)$ is immediate. Let $\mathcal{I}$ be an arbitrary interpretation. We can show $\mathrm{deg}_\mathcal{I}^G(F_1) = \mathrm{deg}_\mathcal{I}^G(F_2)$ by distinguishing all cases for $\mathrm{deg}_\mathcal{I}^G(A), \mathrm{deg}_\mathcal{I}^G(B), \mathrm{deg}_\mathcal{I}^G(C) \in \{\mathbb{Z}^-,\mathbb{Z}^+\}$.
    We demonstrate the $\mathrm{deg}_\mathcal{I}^G(A) \in \mathbb{Z}^-$, $\mathrm{deg}_\mathcal{I}^G(B) \in \mathbb{Z}^-$, and $\mathrm{deg}_\mathcal{I}^G(C) \in \mathbb{Z}^-$. Then $\mathrm{deg}_\mathcal{I}^G(A \x B) = \mathrm{deg}_\mathcal{I}^G(A) - \mathrm{opt}^G(B)$ and $\mathrm{deg}_\mathcal{I}^G(B \x C) = \mathrm{deg}_\mathcal{I}^G(B) - \mathrm{opt}^G(C)$. Thus, $\mathrm{deg}_\mathcal{I}^G(F_1) = \mathrm{deg}_\mathcal{I}^G(A \x B) - \mathrm{opt}^G(C) = \mathrm{deg}_\mathcal{I}^G(A) - \mathrm{opt}^G(B) - \mathrm{opt}^G(C)$. Moreover, $\mathrm{deg}_\mathcal{I}^G(F_2) = \mathrm{deg}_\mathcal{I}^G(A) - \mathrm{opt}^G(B \x C) = \mathrm{deg}_\mathcal{I}^G(A) - (\mathrm{opt}^G(B) + \mathrm{opt}^G(C)) = \mathrm{deg}_\mathcal{I}^G(F_1)$.
\end{proof}

Secondly, it follows directly from the above degree function that negation in our new semantics behaves as desired\footnote{Recall the discussion in Section~\ref{subsect:issues}.}. Crucially, negation acts both on hard- and soft-constraints. Moreover, $F$ and $\neg F$ are never satisfied by the same interpretation, $F$ and $\neg\neg F$ are equivalent, and the degree of $\neg F$ depends only on the degree of $F$.

\begin{lemma}\label{lemma:new-semantics-negation}
    Let $F$ be any GCL-formula, and let $\mathcal{I}$ and $\mathcal{J}$ be interpretations. It holds that 
    \begin{itemize}
        \item $\mathrm{deg}_\mathcal{I}^G(F) \in \mathbb{Z}^+$ iff $\mathrm{deg}_\mathcal{I}^G(\neg F) \in \mathbb{Z}^-$,
        \item $\mathrm{deg}_\mathcal{I}^G(F) = \mathrm{deg}_\mathcal{I}^G(\neg \neg F)$, 
        \item if $\mathrm{deg}_\mathcal{I}^G(F) = \mathrm{deg}_\mathcal{J}^G(F)$ then $\mathrm{deg}_\mathcal{I}^G(\neg F) = \mathrm{deg}_\mathcal{J}^G(\neg F)$.
    \end{itemize}
\end{lemma}

Intuitively, negation in GCL preserves information on preferences by allowing for \emph{degrees of dissatisfaction}. For example, the formula $\neg(a \x b)$ can only be satisfied by the interpretation $\emptyset$. However, we must also inspect the interpretations that do not satisfy the formula: $\{b\}$ will result in a degree of $-2$ while $\{a\}$ and $\{a,b\}$ will result in a degree of $-1$, meaning that $\{b\}$ is more preferable than $\{a\}$ and $\{a,b\}$. This reflects the fact that negation in GCL not only negates truth, as in QCL, but also preferences. Also note that, unlike in PQCL, the implication $F \rightarrow G$ can be defined via $\neg F \lor G$ since the antecedent $F$ is classically satisfied if and only if $\neg F$ is not.

\subsection{Adequacy of $\mathbf{NG}$} 
Using the notation of Section~\ref{subsec:new-game} and the degree-function from ~Section~\ref{subsec:extracting-degree-semantics}, we can show the following result:

\begin{theorem}
The value of $\mathbf{NG}(\mathbf{P}:F,\mathcal{I})$ is $\mathrm{deg}_\mathcal{I}^G(F)$.
\end{theorem}

\begin{proof}
We use the same notation as in the proof of Theorem~\ref{thm:adeqqcl} and proceed by induction on the following two claims (1) $\delta(\mathbf{P}:F) = \mathrm{deg}_\mathcal{I}^G(F)$ and (2) $\delta(\mathbf{O}:F) = -\mathrm{deg}_\mathcal{I}^G(F)$. Most of the cases are similar to the proof of Theorem~\ref{thm:adeqqcl}, so we focus on the cases $\mathbf{P}: F$ that require different reasoning:

$F = G_1 \x G_2$: From the fact that $\delta$ respects $\ll$ for the winning outcomes of both players and the game rule of $\x$, we observe the following facts: First, if the $G_1$-game is winning for \Me, \I go to $G_1$ in the first round. Secondly, if $G_1$ is losing and $G_2$ is winning, \I go to $G_2$. And thirdly, if both games are losing, \I go to $G_1$. Since all outcomes of the $G_2$-games are in $\ll$-relation to all outcomes of the $G_1$-game, we have by Lemma~\ref{lem:longestpathng} for all outcomes $O$:
\[\delta_{\mathbf{P}:F}(O) = \begin{cases}
\delta_{\mathbf{P}:G_1}(O), &\text{ if } O\in  W(\mathbf{P}:G_1),\\
\delta_{\mathbf{P}:G_2}(O)+\mathrm{opt}(G_1), &\text{ if } O\in   W(\mathbf{P}: G_2),\\
\delta_{\mathbf{P}:G_1}(O)-\mathrm{opt}(G_2), &\text{ if } O \in L(\mathbf{P}:G_1).\end{cases}\]
The last case comes from the fact that $O \gg O'$ for all $O' \in \mathcal{O}(\mathbf{P}:G_2)$, Lemma~\ref{lem:longestpathng} and the definition of $\delta$. We now use the inductive hypothesis: in the first case from above, $O({\mathbf{P}:F}) \in  W({\mathbf{P}:G_1})$ and therefore $\delta({\mathbf{P}:F}) = \delta({\mathbf{P}:G_1}) = \dI^G(G_1)$. In the second case, $O({\mathbf{P}:F}) \in W({\mathbf{P}:G_2})$ and therefore $\delta({\mathbf{P}:F}) = \delta({\mathbf{P}:G_2})+\mathrm{opt}(G_1) = \dI^G(G_2)+\mathrm{opt}(G_1)$. Finally, in the third case, $O({\mathbf{P}:F}) \in L({\mathbf{P}:G_1})$ and therefore $\delta({\mathbf{P}:F}) = \delta({\mathbf{P}:G_1})-\mathrm{opt}(G_2) = \dI^G(G_2)-\mathrm{opt}(G_2)$.

$F = \neg G$: The game continues at ${\mathbf{O}:G}$. Therefore, using the inductive hypothesis (2),  $\delta({\mathbf{P}:F}) = \delta({\mathbf{O}:G}) = -\mathrm{deg}_\mathcal{I}^G(G) = \mathrm{deg}_\mathcal{I}^G(F)$.

Cases where \I am in the role of Opponent are similar. For example, let us consider ${\mathbf{O}: G_1 \wedge G_2}$. In the first move \I choose between the two subgames ${\mathbf{O}:G_1}$ and ${\mathbf{O}:G_2}$. \I seek to maximize \My payoff, so \I go to the subgame with $\trianglelefteq$-maximal value. Therefore, using the inductive hypothesis, $\delta({\mathbf{O}:G_1 \wedge G_2}) = \max\{\delta({\mathbf{O}: G_1}), \delta({\mathbf{O}: G_2})\} = \max\{-\dI^G(G_1), -\dI^G( G_2)\} \allowbreak = - \min\{\dI^G(G_1),\dI^G(G_2)\} = -\dI^G(G_1\wedge G_2)$.
\end{proof}

\section{Towards Preferred Model Entailment}
In future work we plan to present a lifting of the GTS for GCL to a \emph{provability game} for preferred model entailment\footnote{In preferred model entailment, a set $T$ of QCL-formulas entails a classical formula $F$ if $F$ is true in all preferred models of $T$.}. This lifting is done in two steps. First, the game is extended to a family of \emph{truth-degree comparison games} \cite{FerLanPav20, AlexandraGoedel} parametrized by $r\in Z$. The rules of this game closely follow the rules of $\mathbf{NG}$ to ensure that \I have a winning strategy in the new game iff \I have a strategy for $\mathbf{NG}$ with payoff $\trianglerighteq r$.

In the second step this game is lifted to a \emph{disjunctive game} \cite{FMGiles,RobHybridLogic}. Intuitively, the two players play all GTS-games over a fixed formula simultaneously over all models. Additionally, \I am allowed to make back-up copies of game states, which \I can return to later. \I win this game iff \I win in at least one back-up copy. We will show that the disjunctive game adequalty models preferred model entailment. Furthermore, \My winning strategies directly correspond to proofs in a cut-free sequent-calculus.
The technique of disjunctive states has already been demonstrated for a number of GTS for different logics~\cite{FMGiles,RobHybridLogic,FerLanPav20, AlexandraGoedel}. The case of GCL is the first to require a two-step lifting and give rise to a deduction/refutation system~\cite{Goranko2019-GORHDS}.

\section{Conclusion}

This paper proposes game semantics for the language of Qualitative Choice Logic (QCL), and thereby show that game-theoretic semantics (GTS) are well-suited for logics such as QCL in which soft- and hard-constraints are expressed in a single language.

On the one hand, we show that the degree-based semantics of QCL can be captured naturally via GTS. 
The notion of optionality, which must be defined a-priori in QCL, arises naturally in our setting as a property of game trees.

On the other hand, we make use of GTS negation to introduce a novel semantics for the language of QCL. We show that this new semantics avoids issues with negation in QCL and Prioritized QCL (PQCL) 
while retaining desirable properties such as associativity of ordered disjunction. 

Regarding future work, we outlined how our game semantics can be lifted to a provability game by which a cut-free sequent calculus can be obtained.  We also plan to examine our new semantics with respect to computational properties, and to investigate how our approach can be adapted to formalisms related to QCL such as other choice logics \cite{BoudjelidaBenferhat2016ccl, BernreiterEtAl2020asp, BernreiterEtAl2021} or the lexicographic logic introduced by Charalambidis, Papadimitriou, Rondogiannis, and Troumpoukis~\cite{CharalambidisEtAl2021}.

 \clearpage
\printbibliography
\end{document}